\documentclass[a4paper,fleqn,usenatbib]{mnras}

\usepackage{mathptmx}
\usepackage[usenames, dvipsnames]{color}

\usepackage{caption}
\usepackage[T1]{fontenc}
\usepackage{ae,aecompl}
\usepackage{multirow}
\usepackage{graphicx}	
\usepackage{savesym}
\usepackage{amsmath}
\savesymbol{iint}
\usepackage{txfonts}
\restoresymbol{TXF}{iint}

\usepackage{amssymb}	

\usepackage[subnum]{cases}
\newcommand{\n}{\nonumber}

\newcommand*\df{\mathop{}\!\mathrm{d}}

\usepackage{wrapfig} 
\usepackage{subfig}

\usepackage[english]{babel}
\usepackage[utf8]{inputenc}

\usepackage{hyperref}

\hypersetup{
    colorlinks=false,
    pdfborder={0 0 0},
}

\title[Cosmic ray ionization in diffuse clouds]{What causes the ionization rates observed in molecular clouds?\\ 
The role of cosmic ray protons and electrons}

\author[V. H. M. Phan et al.]{V. H. M. Phan,$^{1}$\thanks{E-mail: \@}
G. Morlino,$^{2,3,4}$
and S. Gabici$^{1}$
\\
$^{1}$APC, Universit\'e Paris Diderot, CNRS/IN2P3, CEA/Irfu, Observatoire de Paris, Sorbonne Paris Cit\'e, France, \\
$^{2}$Gran Sasso Science Institute, Viale Francesco Crispi 7, 67100 L'Aquila, Italy\\
$^{3}$INFN/Laboratori Nazionali del Gran Sasso, Via G.~Acitelli 22, Assergi (AQ), Italy\\
$^{4}$INAF/Osservatorio Astrofico di Arcetri, L.go E.~Fermi 5, Firenze, Italy
}

\date{Accepted XXX. Received YYY; in original form ZZZ}

\pubyear{2015}

\begin{document}
\label{firstpage}
\pagerange{\pageref{firstpage}--\pageref{lastpage}}
\maketitle

\begin{abstract}
Cosmic rays are usually assumed to be the main ionization agent for the interior of molecular clouds where UV and X-ray photons cannot penetrate. Here we test this hypothesis by limiting ourselves to the case of diffuse clouds and assuming that the average cosmic ray spectrum inside the Galaxy is equal to the one at the position of the Sun as measured by Voyager 1 and AMS-02.
To calculate the cosmic ray spectrum inside the clouds, we solve the one-dimensional transport equation taking into account advection, diffusion and energy losses. While outside the cloud particles diffuse, in its interior they are assumed to gyrate along magnetic field lines because ion-neutral friction is effective in damping all the magnetic turbulence.
We show that ionization losses effectively reduce the CR flux in the cloud interior for energies below $\approx 100$ MeV, especially for electrons, in such a way that the ionization rate decreases by roughly 2 order of magnitude with respect to the case where losses are neglected.
As a consequence, the predicted ionization rate is more than 10 times smaller than the one inferred from the detection of molecular lines.
We discuss the implication of our finding in terms of spatial fluctuation of the Galactic cosmic ray spectra and possible additional sources of low energy cosmic rays.  
\end{abstract}




\section{Introduction}
As molecular clouds (MCs) shield quite effectively both UV photons and X-rays \citep{mckee1989,krolikkallman,silk}, cosmic rays (CRs) seem to be the only capable agents to ionize their interior. It is for this reason that CRs are believed to play an essential role in determining the chemistry \citep{dalgarno2006} and the evolution of these star-forming regions \citep[e.g.,][]{wurster}. Recent observations (\citealt{caselli1998,indriolo}, see \citealt{padovani2009} for a review) have suggested that the CR induced ionization rate decreases for increasing column density of MCs and it varies from around $\approx 10^{-16}$ s$^{-1}$ for diffuse MCs down to $\approx 10^{-17}$ s$^{-1}$ for dense ones. 

The ionization rates measured in MCs are tentatively interpreted as the result of the penetration of ambient CRs into clouds \citep{padovani2009}. Thus, in order to model this process and test this hypothesis one needs to know: {\it i)} the typical spectrum of low energy CRs in the Galaxy, and {\it ii)} the details of the transport process of CRs into MCs.   
Remarkably, the spectra of both proton and electron CRs in the local interstellar medium (ISM) at least down to particle energy of a few MeVs are now known with some confidence, thanks to the recent data collected by the Voyager probe at large distances from the Sun \citep{stone2013,cummings2016}. Whether or not such spectra are the representative of the average Galactic spectra, especially for MeV CRs, is still not clear \citep[this is an old standing issue, see e.g.][]{cesarsky1975}. However, the analysis of gamma rays from MCs \citep[e.g.,][]{yang2014} seems to indicate that at least the spectrum of proton CRs of energy above a few GeV is quite homogeneous in our Galaxy.

Several theoretical estimates of the CR induced ionization rate in MCs have been performed over the years. The first attempts were done by simply extrapolating to low energies the spectra of CRs observed at high energies, without taking into account the effect of CR propagation into clouds \citep[e.g.,][]{hayakawa,spitzer,nath,webber1998}. 
Such estimates provide a value of the CR ionization rate which is known as the {\it Spitzer value} and is equal to $\approx 10^{-17}$ s$^{-1}$. This value is an order of magnitude below the observed data for diffuse clouds, and roughly similar to the value found in dense ones. 

Later works included also a treatment of the transport of CRs into MCs and considered the role of energy losses (mainly ionization) suffered by CRs in dense and neutral environments. A natural starting point is to consider the scenario that maximises the penetration fo CRs into clouds. This was done, most notably, by \citet{padovani2009}, who assumed that CRs penetrate MCs by moving along straight lines. A more realistic description, however, should take into account the fact that the process of CR penetration into MC is highly nonlinear in nature. This is because CRs themselves, as they stream into the cloud, generate magnetic turbulence through streaming instability \citep{wentzel}. The enhanced magnetic turbulence would, in turn, induce an increase in the CR scattering rate onto MHD waves, which regulates their flux into clouds. The exclusion mechanism of CRs from MCs due to this type of self-generated turbulence was first studied in the pioneering works of \citet{skillingstrong}, \citet{cesarskyvolk}, and \citet{morfill}, while recent studies in this direction include the works by \citet{morlinogabici}, \citet{schlickeiser2016}, and \citet{ivlev2}. 
Nevertheless, none of the models including a more thorough treatment of the effect of CR penetration into clouds has been confronted with the available observational data. 

The main goal of this article is to fill this gap and provide the first comparison between the theoretical predictions from detailed models of CR transport and the measured values of the CR ionization rate in MCs, with a focus onto diffuse ones.
We anticipate here the main results obtained in the following: the intensity of CRs in the local ISM as revealed by Voyager measurements is too weak to explain the level of ionization rate observed in clouds. Possible solutions to this problem include the presence of another source of ionization or a non-uniform intensity of low energy CRs throughout the Galaxy.

The paper is organized as follows: in Sec.~\ref{Sec:Model} we describe a model for the penetration of CRs into clouds. The model is then used to derive the spectra of CR protons and electrons inside MCs (Sec.~\ref{Sec:CRspectra}) and to predict the CR ionization rate, which is then compared to available data (Sec.~\ref{Sec:Ionization}). We discuss our results and we conclude in Sec.~\ref{Sec:Conclusion}.

\section{A model for the penetration of cosmic rays into diffuse clouds}
\label{Sec:Model}

The penetration of CRs into diffuse clouds is described by means of a one-dimensional transport model, where CRs are assumed to propagate only along magnetic field lines. This is a good description of CR transport provided that: {\it i)} the propagation of particles {\it across} magnetic field lines can be neglected, and {\it ii)} the spatial scales relevant to the problem are smaller than, or at most comparable to the magnetic field coherence length in the ISM (here we assume $\approx 50-100$ pc). 
Both conditions are believed to be often satisfied and thus this setup was commonly adopted in the past literature to describe the penetration of CRs into MCs \citep[e.g.][]{skillingstrong,cesarskyvolk,morfill,everettzweibel,morlinogabici,schlickeiser2016,ivlev2}.

\begin{figure}
\centering
\includegraphics[width=3.2in]{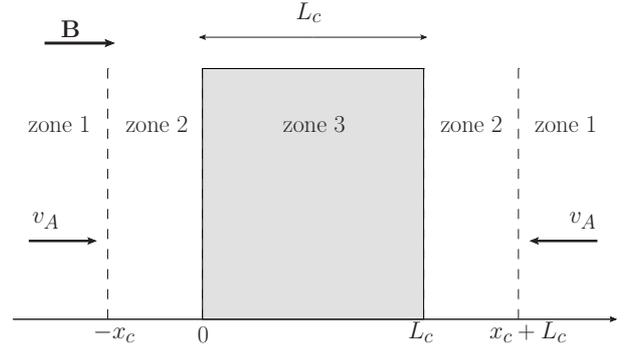}
\caption{Setup of the problem. A cloud of size $L_c$ is embedded in an homogeneous magnetic field of strength $B$ directed along the $x-$axis. The direction of the Alfv\'en speed is also shown. See text for the definition of zone 1, 2,and 3, and of the diffusion scale $x_c$.}
\label{f1}
\end{figure}

In the following, we describe an improved version of the model developed by \citet{morlinogabici}, who considered a diffuse cloud of size $L_c$ and uniform hydrogen density $n_H$ embedded in a spatially homogeneous magnetic field of strength $B$ directed along the $x-$axis (see Fig.~\ref{f1}). 

A spatially uniform field in Zones 1 to 3 (see Fig.~\ref{f1}) is appropriate to describe diffuse clouds, but not dense ones \citep[see the observational results reported in e.g.][]{crutcher} that, for this reason, are not considered in this paper. Note that, for simplicity, the transition from the low density and ionized ISM gas (of density $n_i$) to the dense and neutral cloud environment (of density $n_H \gg n_i$) is taken to be sharp and located at $x = 0$ and $x = L_c$.
\citet{morlinogabici} limited themselves to consider the transport of CR protons only, while here we extend the analysis to include also CR electrons. Moreover, as discussed in the remainder of this Section, we improve the description of the transport of CRs inside the cloud.

In the pioneering papers by \citet{skillingstrong} and \cite{cesarskyvolk} it was suggested that MCs may act as sinks for low energy CRs. This is because low energy CR particles lose very effectively their energy due to severe ionization losses in the dense gas of the cloud.
In steady state, the rate at which CR particles are removed from the cloud due to energy losses has to be balanced by an incoming flux of CR particles entering the cloud \citep{skillingstrong,morlinogabici}.
Therefore, following \citet{morlinogabici}, we consider three regions (see Fig.~\ref{f1}):
\begin{enumerate}
\item{{\bf Zone 1}, located far away from the MC ($x < -x_c$ and $x > x_c+L_c$), where the CR intensity is virtually unaffected by the presence of the cloud. As a consequence, in this zone the CR particle distribution function $f(x,p)$ ($p$ is the particle momentum) is roughly constant in space and equal to the {\it sea} of Galactic CRs $f_0(p)$. The quantity $x_c$ will be defined later.}
\item{{\bf Zone 2}, located immediately outside of the cloud ($-x_c < x < 0$ and $L_c < x < L_c+x_c$). In this zone the CR particle distribution function is significantly affected by the presence of the cloud and is significantly different (i.e. smaller) than $f_0(p)$.}
\item{{\bf Zone 3}, which represents the cloud ($0 < x < L_c$), and where particles suffer energy losses (mainly due to ionization).}
\end{enumerate}

The penetration of CR particles into the cloud is accompanied by the excitation of Alfv\'en waves due to streaming instability \citep{wentzel}. Such instability mainly excites waves propagating in the direction of the streaming of CRs. Therefore, a converging flow of Alfv\'en waves is generated outside of the cloud (see Fig.~\ref{f1}).
Once inside the cloud, Alfv\'en waves are damped very quickly due to ion-neutral friction \citep{zweibelshull}.
For this reason, here we follow \citet{morlinogabici} and we assume that the transport of CRs is diffusive (regulated by the scattering of CRs off Alfv\'en waves) outside of the cloud (Zones 1 and 2), and described by:
\begin{eqnarray}
\label{eq:diffusive}
\frac{\partial f}{\partial t} = \frac{\partial }{\partial x}\left[D\frac{\partial f}{\partial x}\right]-v_A\frac{\partial f}{\partial x} -\frac{1}{p^2}\frac{\partial}{\partial p}\left[\dot{p} p^2 f\right] ~ ,
\end{eqnarray}
while it is ballistic inside the cloud (Zone 3), where Alfv\'en waves are virtually absent (see \citealt{ivlev2} for a discussion on wave transport in clouds). 
In Eq.~\ref{eq:diffusive} above, $f = f(t,x,p)$ is the isotropic part of the CR particle distribution function, which depends on time $t$, position $x$, and particle momentum $p$, $D = D_0(p)$ is the CR diffusion coefficient outside of the cloud (assumed here to be spatially homogeneous), $\dot{p}$ is the rate of momentum loss of CRs (mainly due to interaction between CRs and gas), and $v_A=B^2/\sqrt{4\pi\rho_i}$ is the Alfv\'en speed ($\rho_i$ is the mass density of the ionised gas). Since we assume here that the gas in Zones 1 and 2 is spatially homogeneous, the momentum loss rate $\dot{p}$ is a function of particle momentum only, and the Alfv\'en speed $v_A$ is a constant. Here we search for steady-state solutions and thus we set $\partial f/\partial t = 0$.

To solve the problem, it is convenient to consider separately CRs of high and low energy, with $E_*$ being the energy defining the transition between the two domains (see \citealt{ivlev2} for a similar approach). Following \citet{morlinogabici}, $E_*$ is defined in such a way that particles with energy $E > E_*$ can cross ballistically the cloud without losing a significant fraction of their energy. If $\tau_l$ is the energy loss time of CRs inside the cloud (see Fig.~\ref{f2}), then the energy $E_*$ is obtained by equating $\tau_l$ with the CR ballistic crossing time $\tau_c \sim L_c/\bar{v}_p(E_*)$, where $\bar{v}_p$ is the CR particle velocity averaged over pitch angle (the angle between the particle velocity and the direction of the magnetic field). Obviously, for $E > E_*$ the spatial distribution of CRs inside the cloud is, to a very good approximation, constant. It is important to stress that energy losses play an important role also for particle energies $E > E_*$ (no energy losses in a single cloud crossing), because such CRs are confined in the vicinity of the MC by the converging flow of Alfv\'en waves, and can thus cross and recross the cloud a very large number of times (for a more detailed discussion of this issue the reader is referred to \citealt{morlinogabici}).

\subsection{High energies}
\label{sec:high}

\begin{figure}
\centering
\includegraphics[width=3.42in]{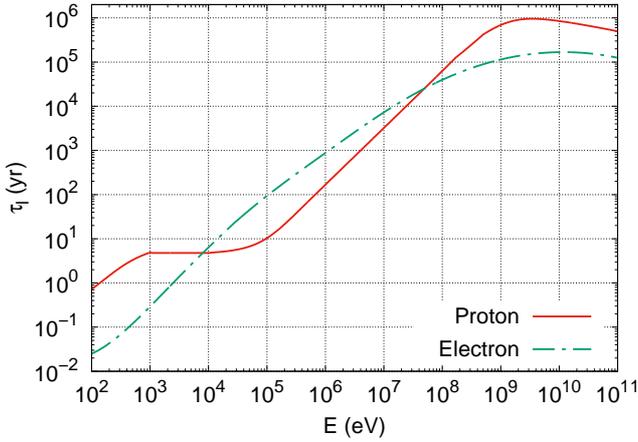}
\caption{Energy loss time for CR protons (red line) and electrons (green line) in a cloud of density $n_H=100$ cm$^{-3}$ (solid lines). The loss times are from \citet{padovani2009}.}\label{f2}
\end{figure}

\citet{morlinogabici} argued that, for $E > E_*$, Eq.~\ref{eq:diffusive} can be also used to describe the transport of CRs inside of the cloud. 
This is because a spatially uniform distribution of CRs can be obtained inside the cloud by assuming a very large value for the particle diffusion coefficient in that region. More quantitatively, the assumption to be made is: $D_c \gg L_c^2/\tau_l$, where $D_c$ is the diffusion coefficient inside the cloud. Under this approximation, Eq.~\ref{eq:diffusive} can be integrated to obtain an expression for $f(x,p)$ outside of the cloud \citep{morlinogabici}:
\begin{eqnarray}
\label{eq:general}
f(x,p)=f_0(p)-\frac{1}{v_Ap^2}e^{\frac{x}{x_c}}\int^{L_c/2}_{0}\frac{\partial }{\partial p}\left[\dot{p}(p)p^2f(y,p)\right]\df y,
\end{eqnarray}
which for $x=0$ or $x=L_c$ reduces to:
\begin{eqnarray}
f_c(p)=f_0(p)-\frac{L_c}{2 v_A p^2} \frac{\partial }{\partial p}\left[\dot{p}(p)p^2f_c(p)\right] ~ . \label{eq:highE}
\end{eqnarray}
where we used the fact that the spatial distribution of CRs is constant inside the cloud. 
In the expression above, $f_c(y,p)\simeq f_c(p)$ is the CR particle distribution function of CRs inside the cloud and $x_c=D_0/v_A$ is a characteristic length that defines the extension of Zone 2 in Fig.~\ref{f1}.


From Eq.~\ref{eq:highE} a semi-analytical expression for $f_c(p)$ can be easily derived, and it reads \citep{morlinogabici}:

\begin{eqnarray}
\label{eq:solutionhigh}
f_c(p)=\frac{2v_A\tau_l(p)}{L_cp^3}\int^{p^{\text{max}}}_p q^3 f_0(q)\exp\left[-\frac{2v_A}{L_c}\int^q_p\frac{\tau_l(k)}{k}\df k\right]\df q,\label{eq:fchighE}
\end{eqnarray}
where we have introduced the loss time inside of the cloud $\tau_l(p)=-p/\dot{p}$. For the energy losses we adopt the same expression used by \cite{padovani2009}. The corresponding energy loss time is also reported in Fig.~\ref{f2} for both protons and electrons.
In deriving Eq.~\ref{eq:fchighE} we implicitly assumed that $\dot{p} \sim 0$ in Zones 1 and 2.
This is a valid assumption for both protons and electrons, because the energy loss time outside of the cloud is much longer than the characteristic dynamical time of the problem, which can be defined as $\sim D_0/v_A^2$ \citep{morlinogabici}.

As said above, Eq.~\ref{eq:fchighE} provides a general solution for spectrum of CRs with energy $E > E_*$, or equivalently, of momentum larger than $p > p_*$. The numerical values for the critical energy $E_*$ and momentum $p_*$ can be found from the expression $\tau_l(p_{*})\simeq 2 L_c/v_p(p_*)$ where $v_p$ is the speed of a particle of momentum $p_*$ (here we set $\bar{v}_p=v_p/2$). 
For a cloud of size $L_c=10$ pc and $n_H=100$ cm$^{-3}$ (or equivalently of column density $N_{\text{H}_2}=n_HL_c=3.1 \times 10^{20}$ cm$^{-2}$), we found $p_{*,p} \sim 75$ MeV/c and $p_{*,e} \sim 0.34$ MeV/c corresponding to a kinetic energy of $E_{*,p} \sim 3.0$ MeV and $E_{*,e} \sim 0.10$ MeV for protons and electrons, respectively. 

\begin{figure*}
\centering
\centerline{
\subfloat[CR protons]
  {\includegraphics[width=3.42in]{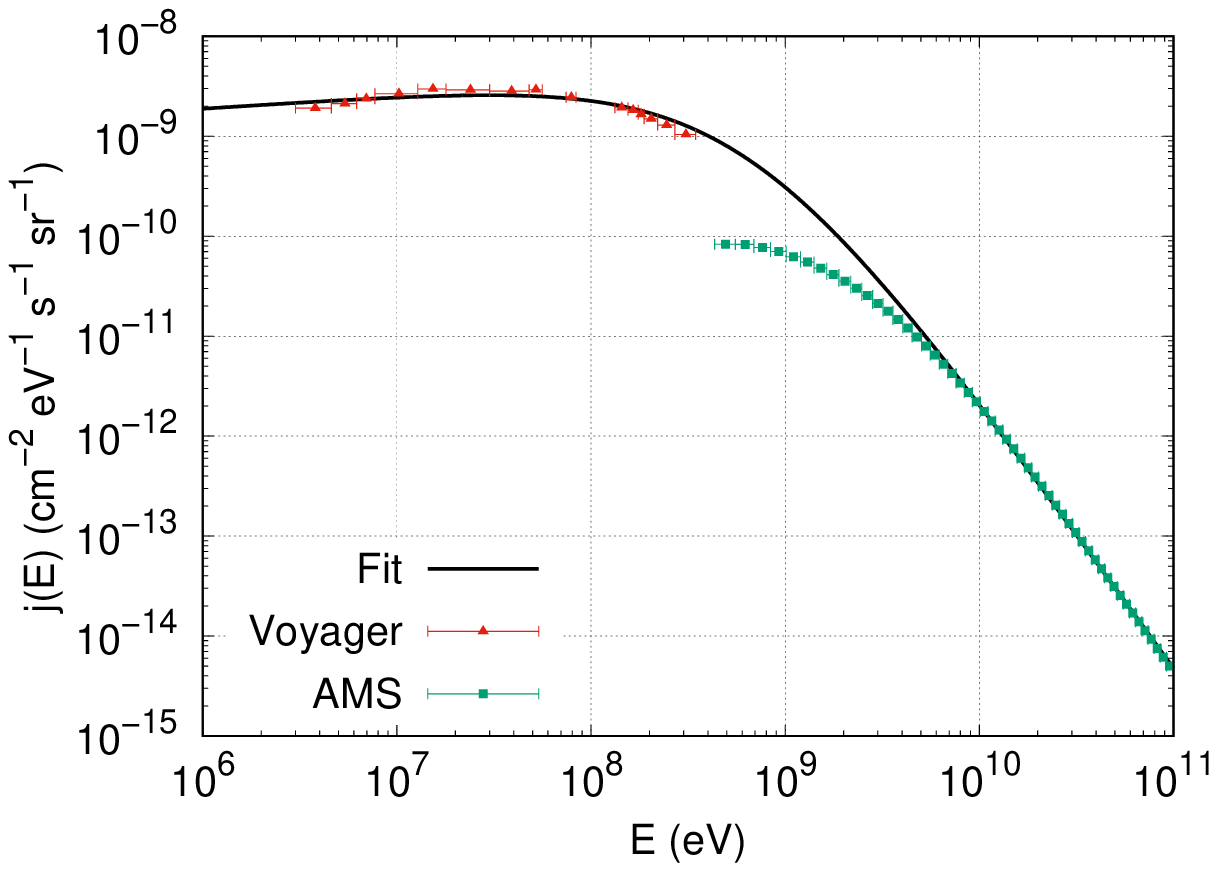}\label{f0-a}}
\subfloat[CR electrons]
  {\includegraphics[width=3.42in]{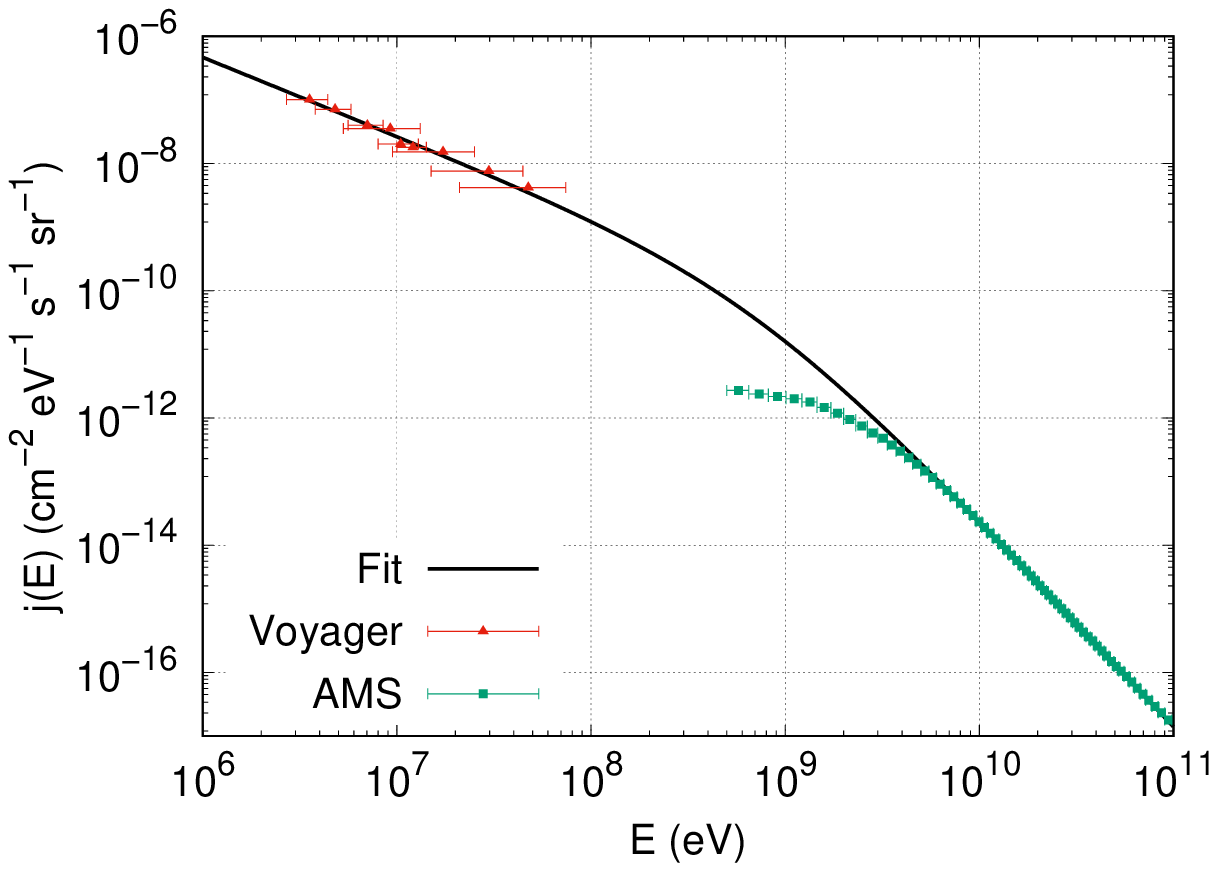}\label{f0-b}}
}  
\caption{Data of the CR intensity for protons (left) and electrons (right) taken from  Voyager 1 \citep{cummings2016} and AMS-02 \citep{AMS2014,AMS2015} compared with the fitted curve used in this work.}\label{f0}
\end{figure*}

\subsection{Low energies}
\label{Sec:LowE}
Particles lose a significant fraction of their energy $E$ in a cloud crossing time $\tau_c$ if $E < E_*$. In this case, the approach described in the previous Section still provides a good description of CR transport outside of the cloud (Zones 1 and 2 in Fig.~\ref{f1}), but might fail inside of the cloud (Zone 3). The reason is that at such low energies the spatial distribution of cosmic rays in Zone 3 is not necessarily constant. 
Thus, in order to describe the transport of CRs inside of the cloud, we will adopt the continuously slowing down approximation as done in \cite{padovani2009}.
This consists in connecting the momentum $p$ of a particle located at a position $x$ inside the cloud to the momentum the particle had when it entered the cloud. We will denote this momentum as $p_{01}$ or $p_{02}$ for particles that entered the cloud from the left and right edge of the cloud, respectively.
Thanks to the symmetry of the problem (the flux of CRs impinging onto the left and right side of the cloud is identical) we can write:
\begin{eqnarray}
f_c(x,p)\df^3 p=\frac{1}{2} \left[ f_b(p_{01})\df^3 p_{01}+f_b(p_{02})\df^3 p_{02}\right].\label{eq6}
\end{eqnarray}
where $f_b(p)$ is the CR particle distribution function at the cloud border, which is assumed to be quite close to an isotropic distribution. 
Eq.~\ref{eq6} can be reduced to:
\begin{eqnarray}
f_c(x,p)=\frac{1}{2} \left[ f_b(p_{01})\frac{p_{01}^2\dot{p}(p_{01})}{p^2\dot{p}(p)}+f_b(p_{02})\frac{p_{02}^2\dot{p}(p_{02})}{p^2\dot{p}(p)} \right],\label{eq:spatial}
\end{eqnarray}
which can be further simplified by noting that $p_{01}=p_0(x,p)$ and $p_{02}=p_0(L_c-x,p)$.

The function $p_0(x,p)$ can be determined by solving the equation: 
\begin{eqnarray} \label{eq:p0}
\displaystyle x=\langle\cos\vartheta\rangle\int^p_{p_0}\frac{v_p}{\dot{p}(p)}dp \approx \frac{1}{2}\int^p_{p_0}\frac{v_p}{\dot{p}(p)}dp,
\end{eqnarray}
where we introduced $\vartheta$ as the particle pitch angle, and we set $\langle\cos\vartheta\rangle\ \simeq 1/2$, as expected for an almost isotropic distribution of particles. Note that, even though deviation from isotropy are expected at low energies (for $E \ll E_*$ one does not expect to have a significant flux of particles out of the cloud), the error introduced by the assumption of CR isotropy is at most a factor of 2 (and most likely significantly less than that, as argued by \citealt{ivlev2}).

At this point we can change coordinate system from $(x,p)$ to $(p_0,p)$, and combine Eq.~\ref{eq:general} with Eq.~\ref{eq:spatial} to obtain:
\begin{eqnarray}
f_b(p)=f_0(p)+\frac{v_p}{4 v_Ap^2\dot{p}(p)}\int^{p_0^{\text{max}}}_{p}\frac{\partial}{\partial p_0}\left[\vphantom{\frac{•}{•}}p_0^2f_b(p_{0})\dot{p}(p_{0})\right]\df p_0,
\end{eqnarray}
where $p_0^{\text{max}}=p_0(L_c,p)$. This can be solved to give:
\begin{eqnarray}
f_b(p)=\frac{f_0(p)+\dfrac{v_p}{4 v_Ap^2 \dot{p}(p)}\left[\vphantom{\frac{•}{•}}\left(p_0^{\text{max}}\right)^2f_b(p_0^{\text{max}})\dot{p}(p_0^{\text{max}})\right]}{1+\dfrac{v_p}{4 v_A} \vphantom{\dfrac{^{\frac{•}{•}}}{•}}},\label{eq:lowE}
\end{eqnarray}
where $p_{0}^{\max}(p)$ is defined by Eq.~(\ref{eq:p0}) with $x=L_c$ and represents the momentum of particles at the border of the cloud that produce particles with momentum $p$ on the other side of the cloud.
Indeed, the expression above still does not give the form of $f_b(p)$ as it requires $f_b(p_0^{\text{max}})$ which, in principle, is unknown. However, the asymptotic behavior would be $f_b(p_0^{max}) \sim f_c(p)$ for sufficiently large particle energies, with $f_c(p)$ given by Eq.~\ref{eq:highE}. 

It is worth mentioning that Eq.~\ref{eq:lowE} is not a formal solution of Eq.~\ref{eq:general} because, in general, one would expect $\langle\cos\vartheta\rangle\ \ne 1/2$. However, we have checked the result obtained from Eq.~\ref{eq:spatial} with the approximate solution obtained by the method of flux balancing (see Section 2 in \citealt{morlinogabici}) and the two results match for particles with $v_p\gg v_A$. 

\section{Cosmic-ray spectra in diffuse clouds}
\label{Sec:CRspectra}
In this Section, we will make use of Eq.~\ref{eq:solutionhigh} and Eq.~\ref{eq:lowE} to determine the spectrum of CR protons and electrons inside a given MC. In order to do so, we will need to specify:
\begin{enumerate}
\item{the spectrum of CR protons $f_0^p(p)$ and electrons $f_0^e(p)$ far away from the cloud (Zone 1 in Fig.~\ref{f1});}
\item{the column density $N_{H_2}$ and the size $L_c$ of the cloud;}
\item{the Alfv\'en speed $v_A$ in the medium outside of the cloud (Zones 1 and 2).}
\end{enumerate}
As pointed out in \citet{morlinogabici}, it is a remarkable fact that the spectrum of CRs inside the cloud does not depend on the CR diffusion coefficient (this quantity does not appear in neither Eq.~\ref{eq:solutionhigh} nor \ref{eq:lowE}).

As a reference case, we will assume that the spectra of CR protons and electrons away from the cloud are identical to those measured by the Voyager 1 probe \citep{stone2013,cummings2016}. This is equivalent to assuming that the spectra measured by Voyager 1 are representative of the entire Galaxy, and not only of the local ISM. We will discuss in Sec.~\ref{Sec:Conclusion} the implications of such an assumption. To describe Voyager 1 data, we fit the intensity of CRs together with the available high energy data from AMS \citep{AMS2014,AMS2015} with a broken power law:
\begin{eqnarray}
j_0(E)=C\left(\frac{E}{1 \text{ MeV}}\right)^\alpha\left(1+\frac{E}{E_{\text{br}}}\right)^{-\beta} \text{ eV}^{-1}\text{cm}^{-2}\text{s}^{-1}\text{sr}^{-1},
\label{fits}
\end{eqnarray}
where $E$ is the particle kinetic energy and $E_{\text{br}}$ is the break energy where the slope changes from $\propto E^{\alpha}$ to $\propto E^{-\beta}$. The fit parameters are presented in Table \ref{tabVoyager} and the corresponding intensities are plotted in Fig.~\ref{f0}.
Even though CR protons and electron spectra have been measured by Voyager only for particle energies larger than few MeV, we extrapolate the fits to lower energies also. As it will be shown in the following, such an extrapolation does not affect at all our results, because particles with energy below few MeV provide a negligible contribution to the ionization rate of clouds.

\begin{table}
\centering
\caption{Parameters of the fits to the CR proton and electron intensity measured by Voyager 1 and AMS-02.}

	\label{tabVoyager}
	\begin{tabular}{lcccr} 
		\hline
		Species & $C$ ($\text{ eV}^{-1}\text{cm}^{-2}\text{s}^{-1} \text{sr}^{-1}$) & $\alpha$ & $\beta$ & $E_{\text{br}}$ (MeV) \\
		\hline
		Proton & $1.882\times 10^{-9}$ & 0.129 & 2.829 & 624.5 \\
		Electron & $4.658\times 10^{-7}$ & -1.236 & 2.033 & 736.2 \\
		\hline
	\end{tabular}
\end{table}

\begin{figure*}
\centering
\centerline{
\subfloat[CR protons]
  {\includegraphics[width=3.42in]{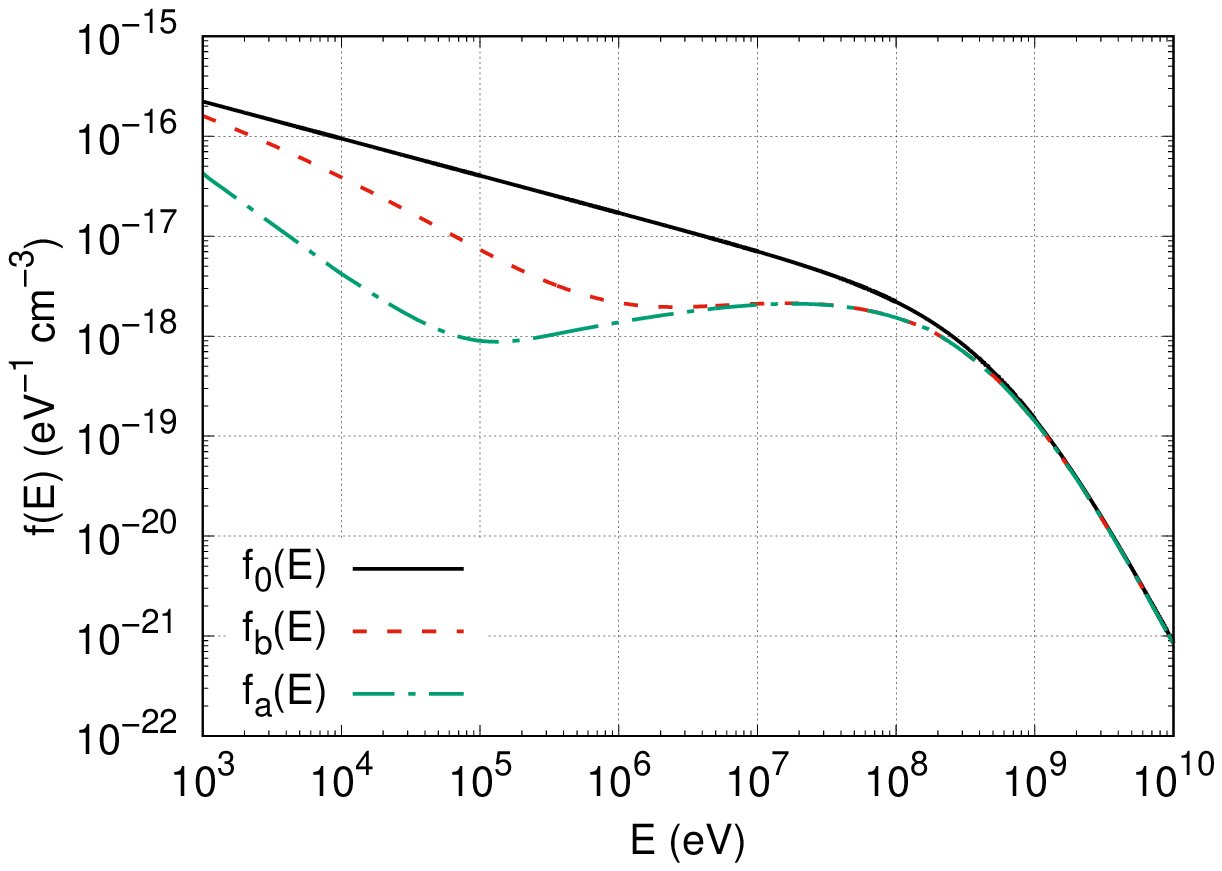}\label{fb-a}}
\subfloat[CR electrons]
  {\includegraphics[width=3.42in]{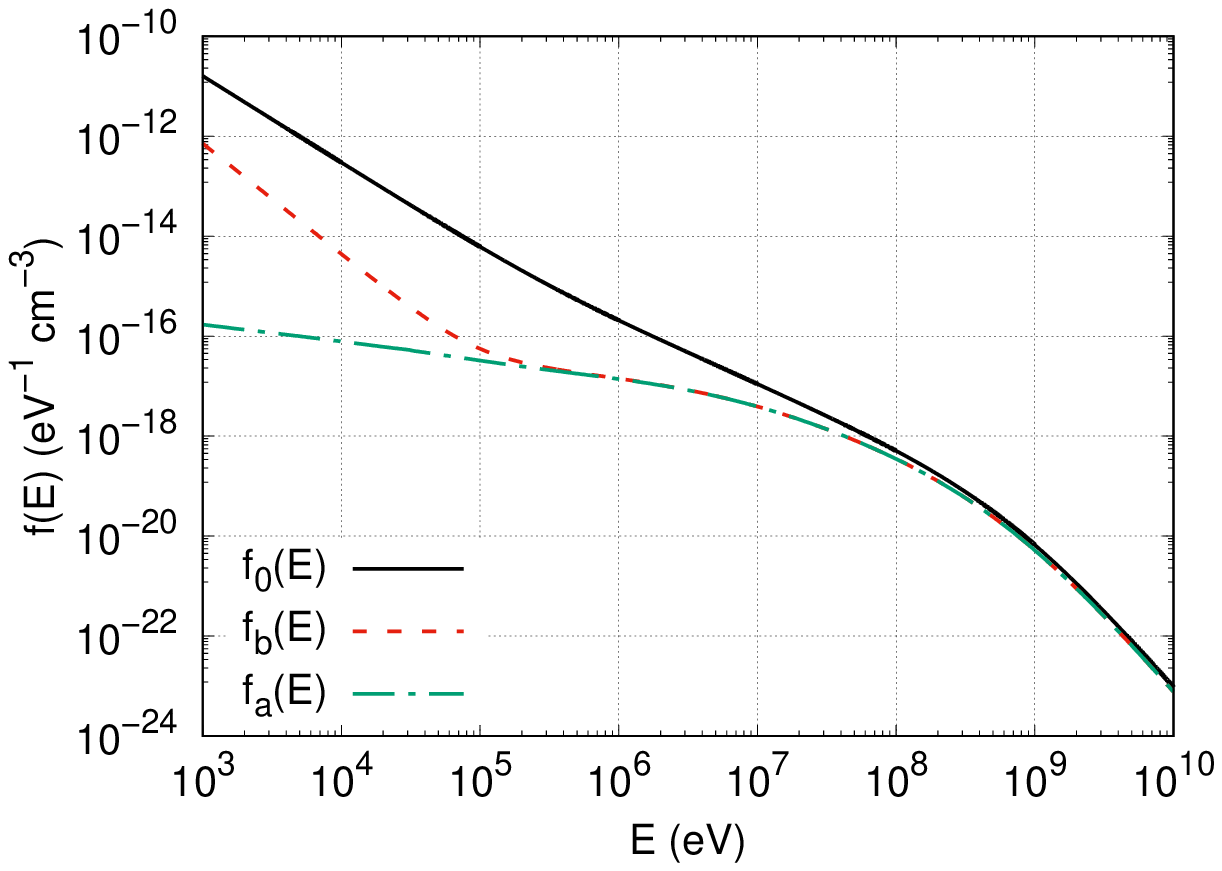}\label{fb-b}}
}  
\caption{CR spectra for a cloud of column density $N_{\text{H}_2}\sim3.1\times 10^{20}$ cm$^{-2}$ (corresponding to typical values of $n_H=100$ cm$^{-3}$ and $L_c=10$ pc). The left and the right figures are respectively spectra of protons and electrons. Also shown with black-solid curves are the ISM spectra given by Eq.~(\ref{fits}).}\label{fb}
\end{figure*}

Results are shown in Fig.~\ref{fb} for both CR protons and electrons, for a cloud of column density $N(\text{H}_2)=3.1\times 10^{20}$ cm$^{-2}$ and for a value of the Alfv\'en speed of $v_A \simeq 200$ km/s. We assume a quite large value for the Alfv\'en speed to maximise the penetration of CRs into clouds. The reason for this assumption will become clear in the following. The three curve represents the spectrum of CRs far away from the cloud $f_0=4\pi j_0(E)/v_p$, the spectrum $f_b$ at the cloud border ($x = 0$ and $x = L_c$), and the spectrum averaged over the cloud volume $f_a$. At large enough energies CRs freely penetrate clouds, and the three spectra coincide. As noticed by \citet{morlinogabici}, this is the case for particles which are not affected by energy losses during their propagation in zones 2 and 3 (see Fig.~\ref{f1}). For the cloud considered in Fig.~\ref{fb} ($N_H \sim 3.1 \times 10^{21}$ cm$^{-3}$) and for a value of the magnetic field of 10 $\mu$G this happens at $E_{loss,p} \sim 39$ MeV for CR protons and $E_{loss,e} \sim 32$ MeV for electrons (see Eq. 7 and related discussion in \citealt{morlinogabici}).
Below these enegies, the proton and electron spectra inside the cloud are suppressed with respect to $f_0$, but in the energy range $E_{*,p} (E_{*,e}) < E < E_{loss,p} (E_{loss,e})$ we still found that $f_b = f_a$. 
$E_{*,p}$ and $E_{*,e}$ have been defined at the end of Sec.~\ref{sec:high}.
This fact can be easily undertood in the following way:
\begin{enumerate}
\item{for proton (electron) energies larger than $E_{loss,p}$ ($E_{loss,e}$) CRs freely penetrate the cloud, so that $f_0 = f_b = f_a$;}
\item{for proton (electron) energies in the range $E_{*,p} < E < E_{loss,p}$ ($E_{*,e} < E < E_{loss,e}$) particles suffer ionisation energy losses, but this happens after they repeatedly cross the cloud. This implies that the CR spatial distribution inside the cloud is uniform, and thus $f_0 \ne f_b = f_a$;}
\item{for proton (electron) energies $E < E_{*,p}$ ($E < E_{*,e}$)) particles lose energy before completing a single crossing of the cloud, which implies that the spatial distribution of CRs inside the cloud is non-uniform, i.e. $f_0 \ne f_b \ne f_a$.}
\end{enumerate}

\begin{figure*}
\centering
\centerline{
\subfloat[CR protons]
  {\includegraphics[width=3.42in]{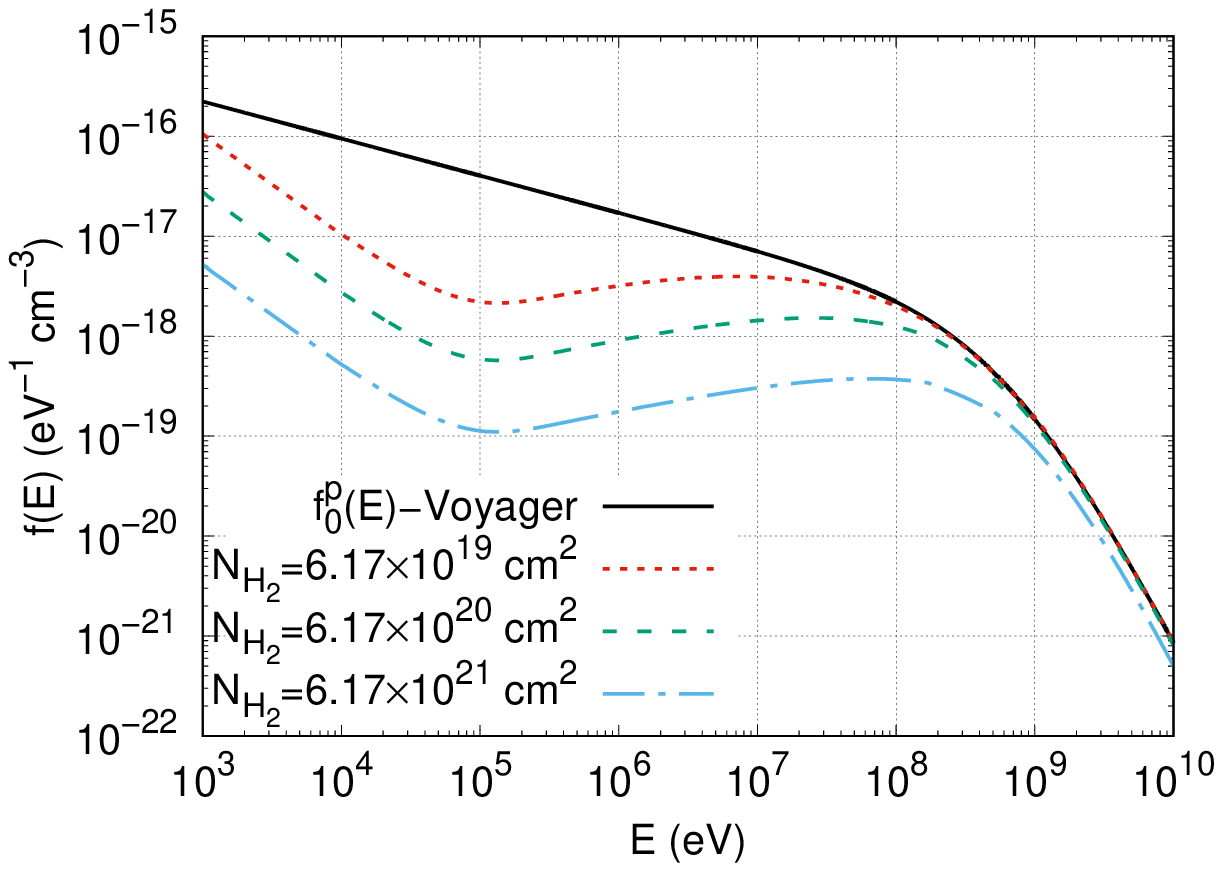}\label{fa-a}}
\subfloat[CR electrons]
  {\includegraphics[width=3.42in]{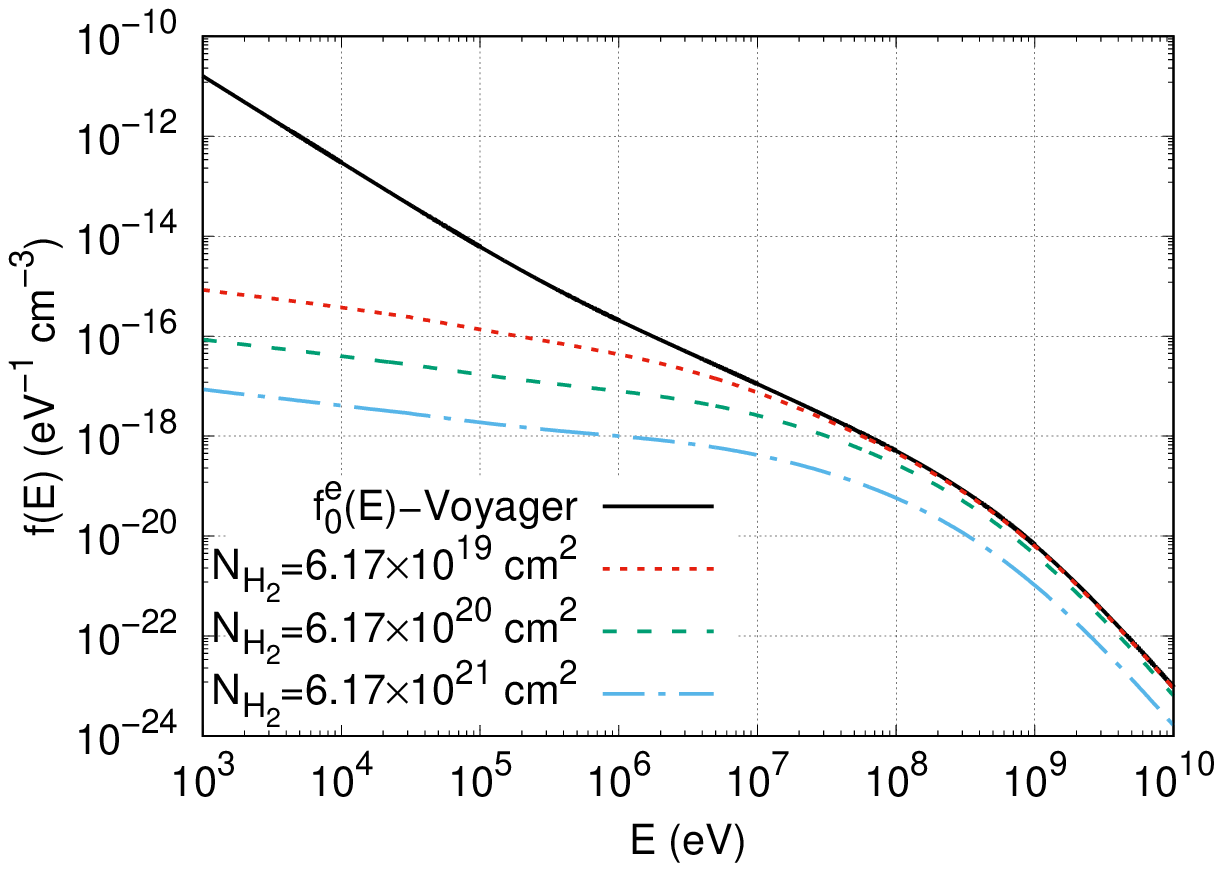}\label{fa-b}}
}  
\caption{Average spectra of CR protons (left panel) and electrons (right panel) inside clouds of different column densities as listed in the labels. The average ISM spectra of Eq.~(\ref{fits}) is also shown with black solid line.}
\label{fa}
\end{figure*}

In Fig.~\ref{fa}, we provide also a few spectra to show how our results depend on the exact value of the column density of the cloud. It is clear from the Figure that the suppression of the CR spectra inside MCs is more pronounced for larger column densities. For very large column densities, approaching $\sim 10^{22}$ cm$^{-2}$, the CR proton and electron spectrum is suppressed with respect to $f_0$ up to quite large energies reaching the GeV domain. 


\section{Ionization rates}
\label{Sec:Ionization}

The CR spectra obtained in the previous Section can now be used to compute the ionization rates $\xi_{\text{H}_2}$ in diffuse clouds.
In the absence of a detailed knowledge of the distribution of the gas of the cloud along the line of sight, we use in the following the spatially averaged spectrum of CRs $f_a$ to compute the ionization rates.
Following \citet{padovani2009} we define the ionization rate of H$_2$ due to protons and electrons as:
\begin{eqnarray}
\xi^p_{\text{H}_2}&=&\int^{E_{\text{max}}}_I f_a(E)v_p\left[1+\phi_p(E)\vphantom{^{\frac{•}{•}}}\right]\sigma^p_{\text{ion}}(E)\df E\n\\
&& +\int^{E_{\text{max}}}_0 f_a(E) v_p \sigma_{\text{ec}}(E)\vphantom{^{\frac{a^b}{•}}}\df E\label{eq:ion}\\
\n\\
\xi^e_{\text{H}_2}&=&\int^{E_{\text{max}}}_I f_a(E)v_p\left[1+\phi_e(E)\vphantom{^{\frac{•}{•}}}\right]\sigma^e_{\text{ion}}(E)\vphantom{^{\frac{a^b}{•}}}\df E
\end{eqnarray}
where $\sigma^p_{\text{ion}}$, $\sigma_{\text{ec}}$, and $\sigma^e_{\text{ion}}$ are the proton ionization cross-section, the electron capture cross-section, and the electron ionization cross-section, respectively. The ionization potential of H$_2$ is $I=15.603$ eV while $\phi_p(E)$ and $\phi_e(E)$ are the average secondary ionization per primary ionization computed as in \cite{krause}. 

\begin{figure*}
\centering
\centerline{
\subfloat[CR protons]
  {\includegraphics[width=3.42in]{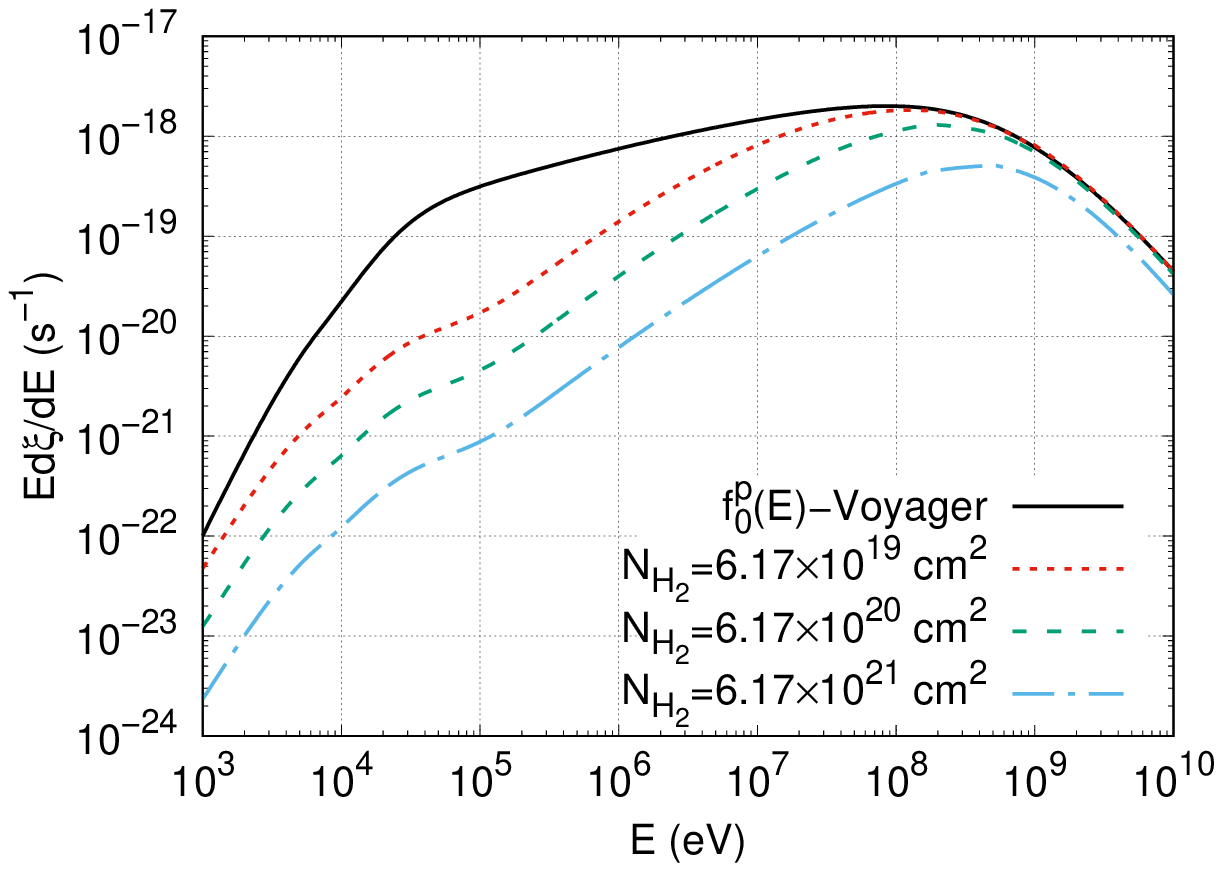}\label{f4a-a}}
\subfloat[CR electrons]
  {\includegraphics[width=3.42in]{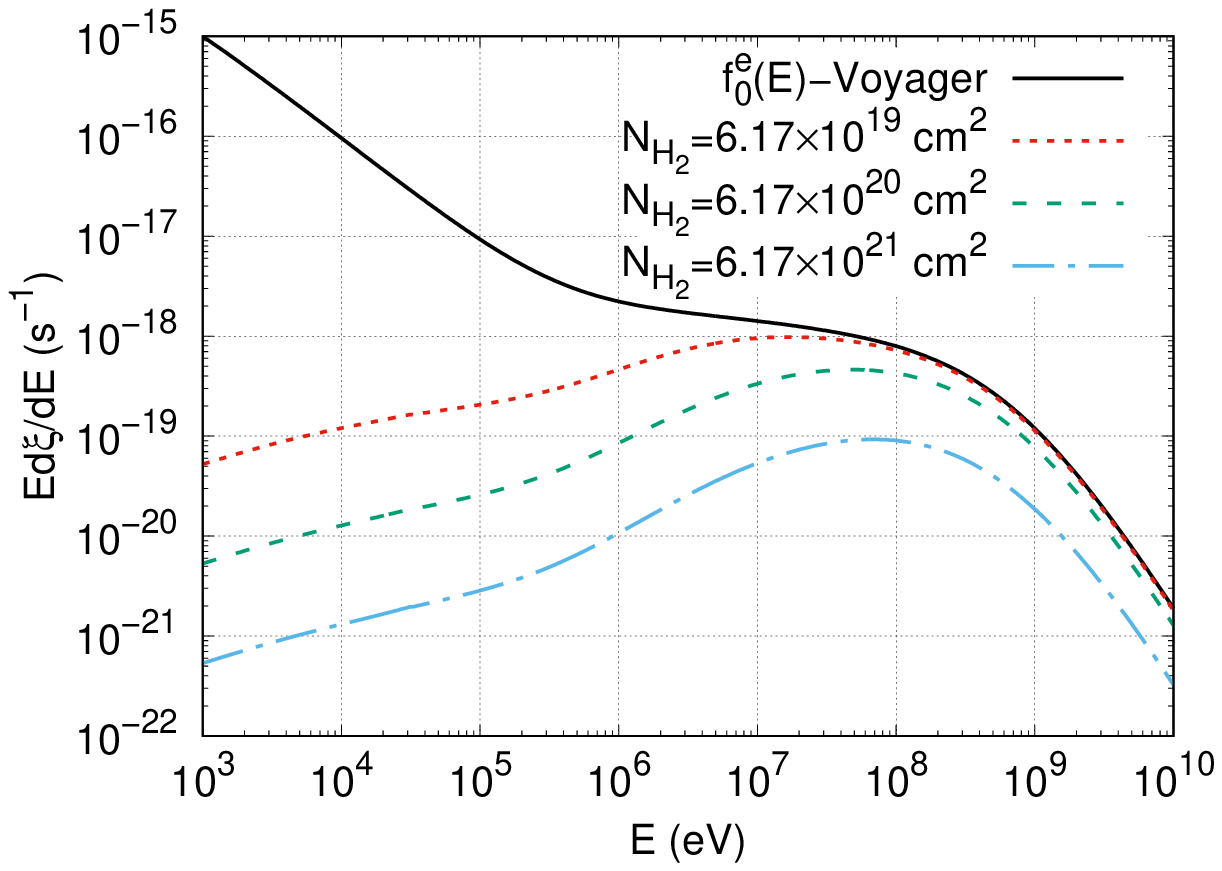}\label{f4a-b}}
}  
\caption{Differential ionization rate of both proton and electron CRs (left and right panel, respectively) at different column densities with the ISM spectra $f_0(E)$ assumed to be that from Voyager and AMS-02 fits. The black curves are the differential ionization rates obtained neglecting propagation and ionization losses into the cloud.}
\label{f4a}
\end{figure*}

Fig.~\ref{f4a} shows the differential contribution to the ionization rate ($E\df\xi_{\text{H}_2}/\df E$) for a few test clouds with column density $6.2\times 10^{19}$, $6.2\times 10^{20}$, and $6.2\times 10^{21}$ cm$^{-2}$ for both protons and electrons. This corresponds to a MC of size $\sim 10$ pc with gas density of $\sim 20$, 200 and 2000 cm$^{-3}$, respectively. The differential ionization rates computed by using the non-propagated Voyager spectra are also shown as black lines. 
These results provide an indication for the range of particle energies that contribute to ionization the most. For the particular case considered here (CRs outside of the cloud have a spectrum equal to that observed by Voyager) it is clear that the differential ionization rate peaks at about $\approx 100$ MeV for both protons and electrons, with a quite weak dependency on the cloud column density.

\begin{figure}
\centering
\includegraphics[width=3.42in]{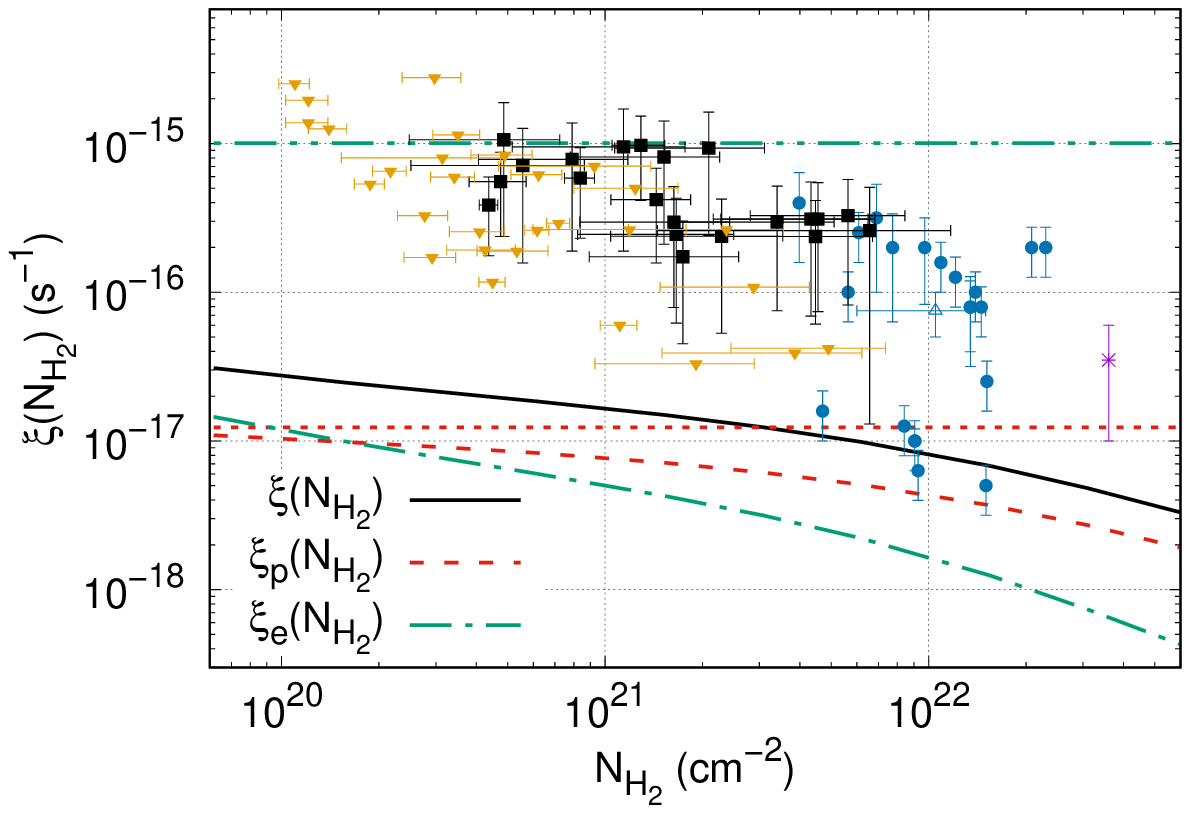}
\caption{Ionization rate derived from Voyager spectra compared to observational data as a function of the column density. The two-dot-dashed line and the dotted line correspond to  the ionization rates of electrons and protons, respectively, neglecting the effects of ionization losses. Data points are from \citet{caselli1998} (blue filled circles), \citet{williams1998} (blue empty triangle), \citet{maret2007} (purple asterisk), and \citet{indriolo} (black filled squares are data points while yellow filled inverted triangles are upper limits).}\label{f4b}
\end{figure}

The dependence of the ionization rate with respect to the cloud column density predicted from our method are shown in Fig.~\ref{f4b}, together with the observational data taken from \citet{caselli1998}, \citet{williams1998}, \citet{maret2007}, and \citet{indriolo}. The ionization rates for protons and electrons ($\xi_{p}(\text{H}_2)$ and $\xi_{e}(\text{H}_2)$, respectively) are plotted together with the total ionization rate, defined as $\xi(\text{H}_2)=\eta\xi_p(\text{H}_2)+\xi_{e}(\text{H}_2)$ where the factor $\eta\simeq 1.5$ accounts for the contribution to the ionization rate from CR heavy nuclei \citep{padovani2009}. 
It is evident from the Figure that the predicted ionization rate fails to fit data, being too small by a factor of several tens at the characteristic column density of diffuse clouds ($N_{H_2} \sim 10^{21}$ cm$^{-2}$).
It seems, then, that the intensity of CRs measured in the local ISM is by far too weak to explain the ionization rates observed in MCs. A discussion of this issue and possible solutions to such a large discrepancy will be provided in the final Section of this paper. It has to be noticed, however, that the predictions presented in Fig.~\ref{f4b} are consistent with the upper limits on the ionization rate measured for a number of clouds (yellow data points).

The range of column densities considered in Fig.~\ref{f4b} encompasses the typical values of both diffuse and dense clouds (a transition between the two regimes can be somewhat arbitrarily set at $N_H \approx {\rm few}~ 10^{21}$ cm$^{-2}$ \citealt{snow}), while the model presented in this paper applies to diffuse clouds only.
The propagation of CRs through large column densities of molecular gas may differ from the description provided in this paper mainly because large column densities are encountered in the presence of dense clumps, where the assumption of a spatially homogeneous distribution of gas density and magnetic field are no longer valid.
The presence of clumps may affect CR propagation mainly in two ways:
\begin{enumerate}
\item{\textit{Magnetic mirroring}: the value of the magnetic field cannot be assumed to be spatially homogeneous in clumps, where it is known to correlate with gas density \citep{crutcher}. The presence of a stronger magnetic field in clumps may induce magnetic mirroring of CRs, as investigated in \citet{padovani2}. This fact would lead to a suppression of the CR  intensity and thus also of the ionization rate. This would further increase the discrepancy between model predictions and data;}
\item{\textit{Particle losses}: very dense clumps may act as sinks for CR particles. This happens when the energy losses are so effective to prevent CR particles to cross the clump over a time-scale shorter than the energy loss time. Such a scenario was investigated by \citet{ivlev2}. Under these circumstances, a larger suppression of the CR intensity inside MCs is expected (energy losses are on average more intense), and this would also increase the discrepancy between data and predictions.}
\end{enumerate}



\section{Discussion and conclusions}
\label{Sec:Conclusion}

The main result of this paper can be summarized as follows: if the CR spectra measured in the local ISM by the Voyager 1 probe are characteristic of the entire ISM, then the ionization rates measured inside MCs are not due to the penetration of such background CRs into these objects, and another source of ionization has to be found. This is a quite puzzling result, which necessarily calls for further studies. 
Several possibilities can be envisaged in order to explain the discrepancy between model predictions and observations. A non-exhaustive list includes:
\begin{enumerate}
\item{\textit{Better description of the transition between diffuse and dense media:} at present, all the available models aimed at describing the penetration of CRs into MCs rely on the assumption of a quite sharp transition between a diluted and ionized medium, and a dense and neutral one. A more accurate description should consider a more gradual transition between these two different phases of the ISM. However, we recall that the simple flux-balance argument mentioned in Sec.~\ref{Sec:Model} and discussed in great detail in \citet{morlinogabici} would most likely hold also in this scenario. It seems thus unlikely that a more accurate modeling could result in a prediction of ionization rates more than one order of magnitude larger than that presented here (as required to fit data);}
\item{\textit{Inhomogeneous distribution of ionizing CRs in the ISM}: the assumption of an uniform distribution of CRs permeating the entire ISM could be incorrect. Fluctuations in the CR intensity are indeed expected to exist, due for example to the discrete nature of CR sources \citep[see for example][and references therein]{GabiciMontmerle2015}. However, gamma-ray observations of MCs suggests that such fluctuations are not that pronounced for CR protons in the GeV energy domain \citep{yang2014}. Thus, fluctuations of different amplitude should be invoked for MeV and GeV particles;}
\item{\textit{CR sources inside clouds}: the ionizing particles could be accelerated locally by CR accelerators residing inside MCs. Obvious candidate could be protostars, which might accelerate MeV CRs, as proposed by \citet{padovani2015,padovani2016}; }
\item{\textit{The return of the CR carrot?} The existence of an unseen component of low energy CRs, called {\it carrot}, was proposed a long time ago by \citet{meneguzzi-audouze-reeves-1971} in order to enhance the spallative generation of $^7$Li, which at that time was problematic. Voyager data strongly constrain such a component, that should become dominant below particle energies of few MeV (the energy of the lowest data points from Voyager). Such a low energy component could also enhance the ionization rate, as recently proposed by \citet{cummings2016}. }
\end{enumerate}

Further investigations are needed in order to test these hypothesis and reach a better understanding of ionization of MCs.

\section*{Acknowledgements}
This project has received funding from the European Union’s Horizon 2020 research and innovation programme under the Marie Skłodowska-Curie grant agreement No 665850.


\label{lastpage}
\end{document}